# Element priors and target support shape chemical transfer in materials graph networks

Ran Zhao[1], Kangming Li[1*]

*[1]Physical Sciences and Engineering Division (PSE), King Abdullah University of Science and Technology (KAUST), Thuwal 23955-6900, Kingdom of Saudi Arabia*

**Correspondence and requests for materials should be addressed to K.L. (email: kangming.li@kaust.edu.sa).*

## Abstract

Materials graph neural networks must often transfer to chemical regions weakly represented in training data. Such transfer can rely on predefined relations among elements or supervised evidence from target-containing structures, but these pathways are usually entangled. Here, held-out-element splits and incremental target support separate their roles. Without target-containing training structures, formation-energy errors depend strongly on the element representation, particularly for H, O and F. Matched perturbations show that representation-induced sharing matters beyond input dimension or numerical form, while a label-free similarity-graph prior reduces selected zero-shot errors. Adding a few target-containing structures sharply lowers errors and contracts differences among one-hot, k-hot and continuous inputs across ALIGNN and CGCNN. Calibration explains only part of this recovery, and freezing the initial element projection preserves most gains in five of six ALIGNN splits. Target support therefore shifts chemical transfer from reliance on static element relations toward learning from target-containing environments.

## INTRODUCTION

Machine learning has become a central tool for predicting crystal properties across large chemical and structural spaces.[1–3] This progress has been enabled by high-throughput density-functional-theory databases, reusable materials descriptors, and standardized benchmarks such as Materials Project, JARVIS, Matbench, and related materials-informatics frameworks.[4–7] Graph neural networks have further advanced this direction by representing crystals as atomistic graphs and learning from local atomic environments rather than only from hand-crafted composition descriptors.[8–10] With the growth of large databases and pretrained atomistic models, a central question is no longer only whether a model can fit a benchmark, but when it can transfer reliably to chemical regions that are weakly represented during training.[3,6,10–14]

Out-of-distribution (OOD) generalization in materials learning is heterogeneous. Splits that are based on temporal distribution, structural symmetry, property range, or chemistry type do not test the same failure mode. Random or in-distribution validation can underestimate prospective deployment error because training and test structures often overlap in composition, structure and property distributions.[6,10] Recent studies have shown sharp degradation under temporal drift, sparse target regions, structure-based OOD splits and extrapolation beyond the property or structure range represented in training data.[15–20] At the same time, not every nominal OOD split is equally difficult: for instance, a prior OOD study shows that many held-out-element tasks are predicted surprisingly well, while it is hard to generalize to H, O and F if these elements are absent in training data.[21]

The uneven difficulty of these LOEO tasks raises a more specific question: when a target element is absent from the training compounds, what allows a crystal GNN to make predictions for that element, and why does success vary so strongly across elements? A crystal GNN does not encounter a held-out element without prior element-level information. Its node features may already encode identity, periodic-table position or physicochemical attributes. Its node features already place elements in relations defined by identity, periodic-table coordinates or physicochemical attributes. We refer to these relations, supplied before task-specific training, as element priors.

Once labeled target-containing structures become available, however, the model can also learn directly from the compound environments in which the target element occurs. The central question is therefore how dependence on the initial element representation changes as direct target support is introduced. Here, zero-shot LOEO denotes training without labeled target-containing structures, while label-free elemental metadata remain available.

Element priors organize this indirect route in different ways. One-hot inputs distinguish identity without encoding graded relations between elements; atomic-number and period/group features impose low-dimensional periodic-table orderings; continuous physicochemical descriptors relate elements through shared attributes; and CGCNN-style (k-hot) descriptors create chemically motivated but hard sharing boundaries[8]. Learned embeddings and attention-based composition models further show that element relations can be organized into representations that improve prediction.[22,23] The element input is therefore an inductive bias for chemical transfer rather than a neutral implementation detail.

Related OOD studies have examined architectures, benchmark construction, dataset scale, feature selection, transfer learning, active data acquisition and representations of complete compounds or crystal structures.[6,10,18–21,23–25] Most directly, physicochemical element features have been shown to improve formation-energy prediction for compounds containing unseen elements.[26] What remains unresolved is how the available route for transfer changes as target support appears. Without labeled target-containing compounds, the held-out element can be connected to observed chemistry only through the element identities, periodic-table coordinates or physicochemical attributes encoded in its node features. Once target-containing compounds are added, the model can also learn directly from their local chemical environments. We therefore ask how strongly predictions depend on these predefined element relations, how quickly that dependence changes with target support and what target-supported retraining learns. We address these questions through matched prior comparisons, perturbation controls, support-response curves, output calibration and frozen-stem diagnostics.

We find that zero-shot LOEO performance is strongly element- and representation-dependent, but this dependence contracts rapidly after a small number of target-containing structures are introduced. Perturbation and similarity-graph experiments show that the organization of element relations matters at zero support, whereas calibration and frozen-stem controls indicate that target-supported gains extend beyond simple output correction or relearning of the initial element projection.

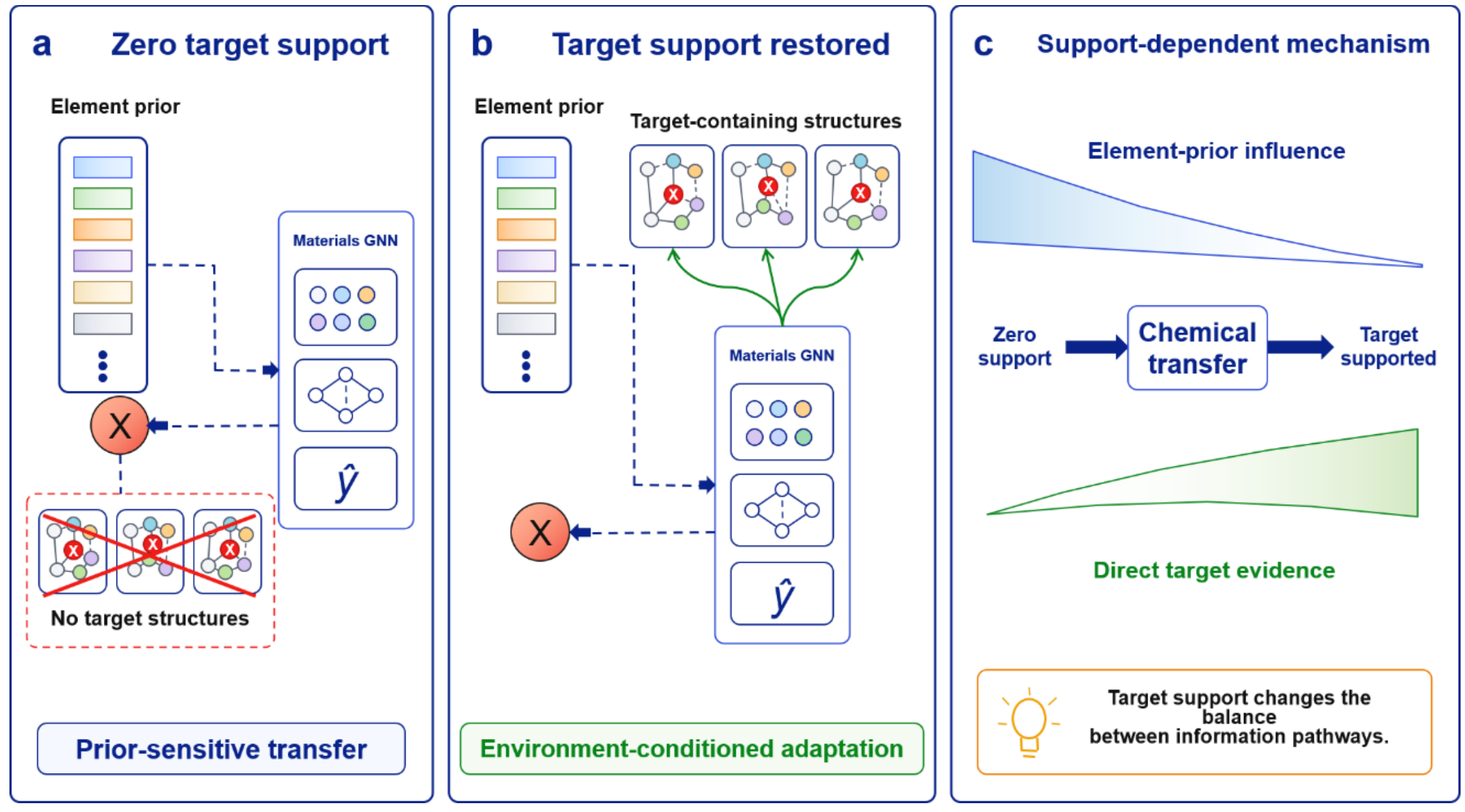

Fig. 1 | Experimental framework for element priors and target support. (a) Zero-target-support LOEO removes all target-containing structures from training; the target element X is represented through the node-feature prior supplied to the materials GNN. (b) Restoring target-containing structures adds labeled compound environments to training while retaining the same model pipeline. (c) Schematic of the two information pathways considered across the support axis. The widths in panel c are conceptual and do not represent fitted functions or measured magnitudes.

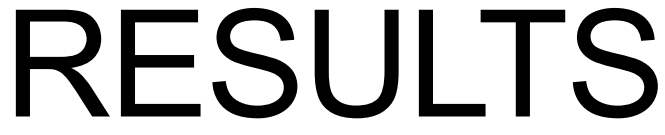

### Experimental logic

We organize the experiments around target support rather than treating LOEO as a single benchmark. The six target elements were chosen to span the task heterogeneity reported previously[21]: Pt, I and B are comparatively lower-error LOEO cases, whereas H, O and F are difficult cases when omitted from training. Including both lower- and higher-error LOEO cases allows us to compare representation and support effects across a range of initial zero-shot difficulty, although the elements are not matched controls. At n = 0, all compounds containing the target element are removed from training. We then add n = 2, 4, 10, 100 or 1000 target-containing structures and ask how prediction error and representation dependence change. ALIGNN provides the primary analysis, and a targeted CGCNN replication tests the cross-architecture trend.

To vary the information available along the zero-support route, we compare seven element inputs. Fixed-zero removes element distinctions and retains only structural geometry; one-hot encodes categorical identity; z and zpg encode atomic number alone or together with period and group; k-hot uses the binned physicochemical descriptor adopted by CGCNN; continuous represents the corresponding attributes as numerical values; and the label-informed reference initializes the element-projection weights from a random-split ALIGNN model. These transferred weights are frozen, while the subsequent interaction and readout layers are trained on each LOEO task. These inputs are treated as distinct inductive biases rather than a scale of increasing chemical information. Dimensions and preprocessing for the tabulated feature inputs are specified in Methods.

The design then probes both sides of this transition. Matched prior comparisons and perturbations examine representation-induced element sharing without target labels; calibration and frozen-stem controls examine what target-supported retraining learns beyond output correction and the initial element projection. Support structures are drawn from the original held-out pool, and the remaining target-containing structures form a disjoint evaluation subset. Table 1 summarizes the experimental logic. Dataset, model, training, perturbation and adaptation procedures are specified in Methods; experiment-specific replicate assignments and elemental-reference sample counts are summarized in Supplementary Tables S2 and S4.

Table 1 | Experimental regimes used to test whether the target-support gain can be attributed to output correction, updates to the element stem, or exposure to target-containing compound environments.

| Regime | Split or experiment | Interpretation |
|---|---|---|
| Zero-shot LOEO | Leave-Pt/I/B/H/O/F-Out | Target-property labels are absent, making input-level element relations especially influential. |
| Target support | Addition of 2, 4, 10, 100 or 1000 target-support structures to the training set | Target-support structures are drawn from the original held-out pool and excluded from the disjoint evaluation subset. |

| Mechanistic target-support controls | Residual calibration; freeze element stem; single-element reference structures | Separates output correction, element-stem adaptation and information from target-containing compound environments. |
| --- | --- | --- |

## Zero-target support exposes target-dependent representation sensitivity

Previous work[26] has shown that physicochemical element features can improve formation-energy prediction for compounds containing elements excluded from training. Here, rather than re-establishing whether chemical metadata can help, we ask how different element representations organize the relation between a held-out element and the chemistry observed during training. We compare the representations under the same ALIGNN architecture and LOEO splits, treating fixed-zero and one-hot as two boundary cases. Fixed-zero removes element distinctions entirely, whereas one-hot preserves categorical identity without encoding a graded relation between the held-out element and seen elements.

In LOEO, the one-hot coordinate assigned only to the held-out element is not constrained by target-element training observations. Its projected representation may therefore be poorly calibrated, while fixed-zero removes this unconstrained identity-specific component. Table 2 compares these boundary cases with the highest-error nonzero representations, and Fig. 2 presents the complete landscape across fixed-zero, one-hot, z, zpg, k-hot, continuous and label-informed inputs.

Table 2 | Boundary comparison between fixed-zero and high-error nonzero representations across LOEO splits. Values are mean ± s.d. over three training seeds.

| Task | Fixed-zero MAE (eV/atom) | Worst nonzero prior | | Second worst nonzero prior | |
| --- | --- | --- | --- | --- | --- |
| | | prior | MAE (eV/atom) | prior | MAE (eV/atom) |
| Pt | 0.393 ± 0.003 | one-hot | 0.373 ± 0.008 | z | 0.287 ± 0.003 |
| I | 0.491 ± 0.003 | one-hot | **0.976 ± 0.063** | z | 0.362 ± 0.005 |
| B | 0.837 ± 0.033 | one-hot | **1.237 ± 0.123** | zpg | 0.756 ± 0.033 |
| H | 2.639 ± 0.355 | one-hot | **2.763 ± 0.257** | continuous | 1.876 ± 0.353 |
| O | 0.923 ± 0.010 | one-hot | **2.829 ± 0.056** | k-hot | 0.751 ± 0.059 |
| F | 1.046 ± 0.025 | one-hot | **2.825 ± 0.161** | k-hot | 1.203 ± 0.024 |

The fixed-zero results are consistent with previous evidence that element-level information can improve prediction for unseen elements: in every LOEO task, at least one structured representation achieves substantially lower error than the geometry-only baseline. The more informative result, however, is that explicit element information is not uniformly beneficial. One-hot is worse than fixed-zero in five of the six tasks, and the relative performance of atomic-number, periodic-table, discretized and continuous representations varies across held-out elements. Performance therefore does not improve monotonically with the amount or apparent chemical detail of the input.

These comparisons shift the focus from whether chemical information is available to how the input representation organizes similarity between the held-out element and the elements observed during training. A useful zero-shot representation must provide relations that support transfer for the target property, rather than simply contain more chemical attributes.

We use the k-hot representation inherited by ALIGNN as a chemically motivated reference[8,9,27–31]. Figure 2 compares k-hot with the other tested inputs under zero target support and shows how strongly the resulting prediction error depends on both the held-out element and the element-similarity structure imposed by the representation.

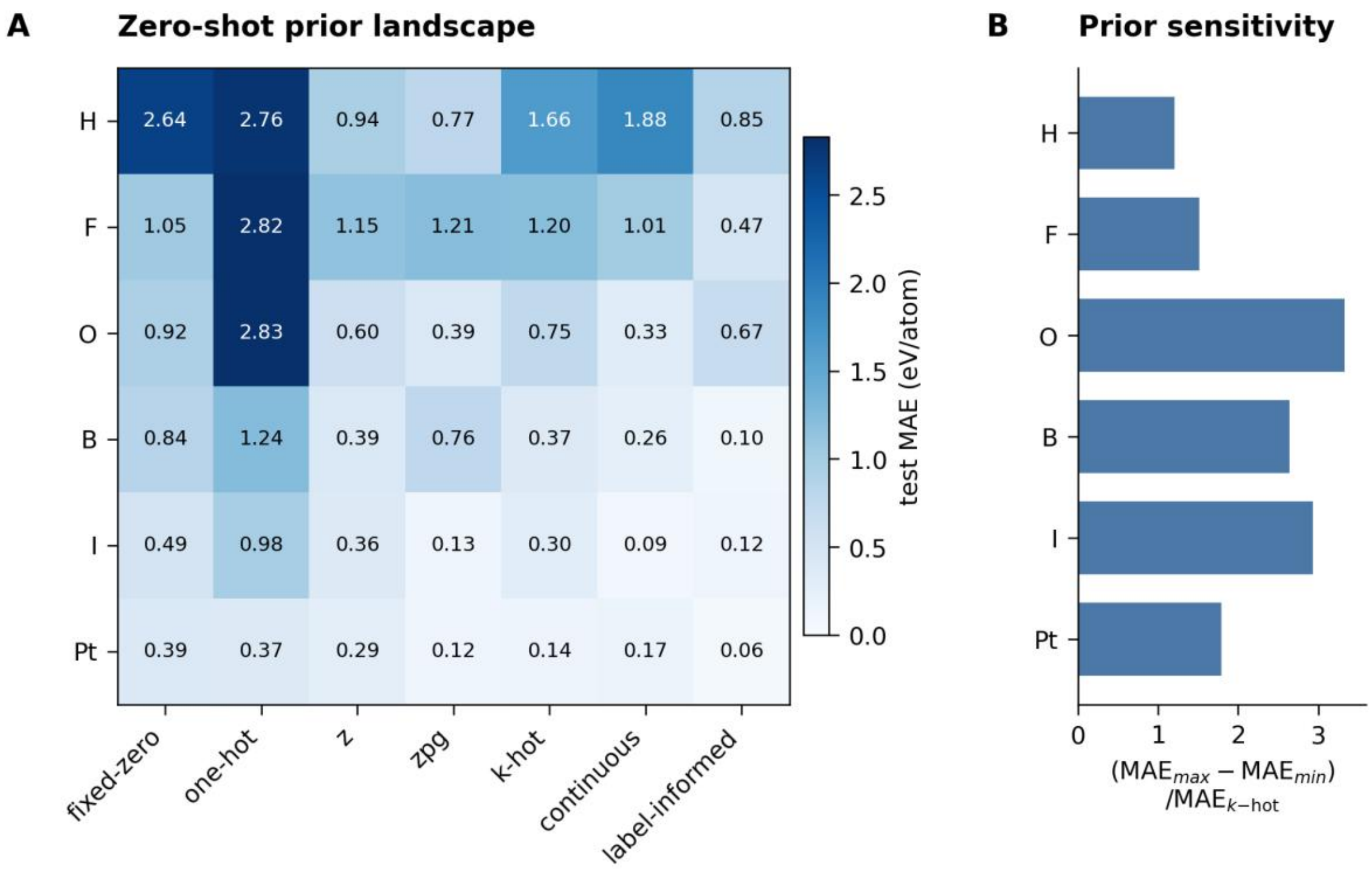

Fig. 2 | Zero-shot LOEO errors and normalized prior sensitivity. (a) Mean test MAE (eV/atom) over three training seeds for Leave-H/F/O/B/I/Pt-Out using fixed-zero, one-hot, z, zpg, k-hot, continuous and label-informed inputs. Cell labels report the mean MAE. (b) Normalized prior sensitivity for each split, defined as $(MAE_{max} - MAE_{min})/MAE_{k\text{-}hot}$ over one-hot, z, zpg, k-hot and continuous inputs.

Pt, I and B have comparatively lower absolute errors, but their preferred representations differ. Continuous features perform well for I, whereas period/group coordinates are already effective for Pt; other inputs produce larger errors in the same tasks. Low absolute error therefore does not imply that the element representation is irrelevant.

H, O and F combine high absolute zero-shot errors with unstable prior rankings. One-hot performs particularly poorly for all three, while no structured prior is consistently best across these tasks. One-hot yields the highest errors among the explicit element inputs for all three tasks, while the structured priors differ substantially and no tested zero-shot representation resolves all three. These splits therefore provide the clearest setting in which to follow how representation dependence changes when target support is introduced.

These zero-shot comparisons establish divergence among representations, but not whether that divergence persists after target-containing environments become available. We therefore next add target-support structures and measure both the error response at a fixed prior and the convergence among different priors.

### Target support rapidly reduces error and differences among priors

At $n = 0$, the initial representation is the only predefined bridge between a held-out element and seen chemistry. To observe how this constraint changes, we restore $n = 2, 4, 10, 100$ or $1000$ target-containing structures while keeping the target-element evaluation subset disjoint. The main support-response curves hold the k-hot prior fixed, so their changes isolate the effect of target support rather than a change of representation.

With the prior held fixed, most of the measured improvement occurs within the first ten target-support structures. Target support reduces error for all six held-out elements, with the largest early gains for H and F. Their MAEs fall by more than 75% after only two target-support structures and continue to decrease at $n = 4$. By $n = 10$, all six tasks reach MAEs below 0.2 eV/atom. For H, O and F, these ten structures recover 84.9–94.7% of the improvement obtained at $n = 1000$. The steep early decline shows that the zero-shot error is highly responsive to limited direct support rather than requiring a large target-domain dataset.

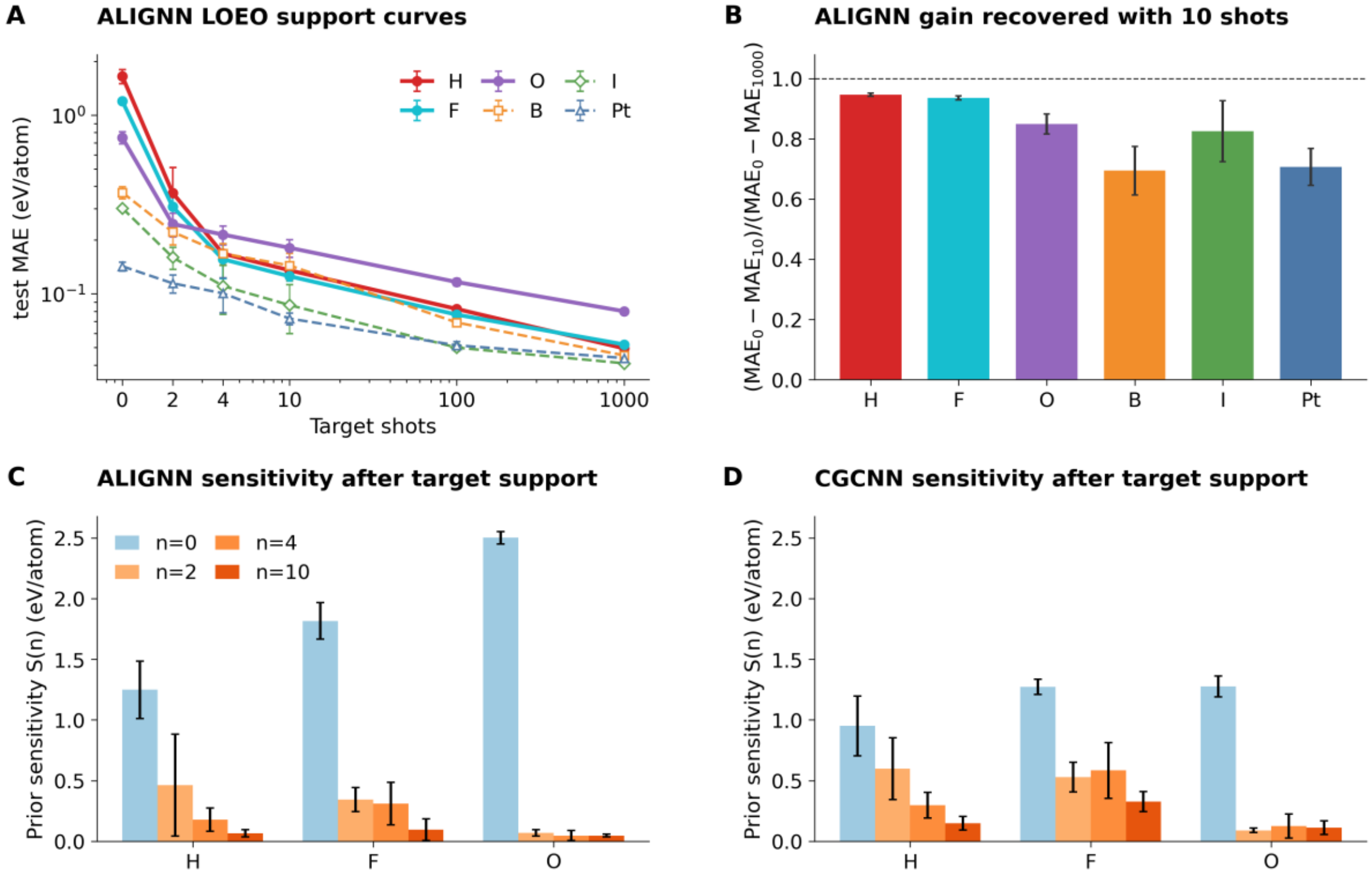

Fig. 3 | Target-support response and prior sensitivity. (a) ALIGNN test MAE on the disjoint target-element evaluation set for Pt, I, B, H, O and F at n = 0, 2, 4, 10, 100 and 1000 target-support structures, using the k-hot prior. (b) Fraction of the 1000-structure improvement measured at n = 10, calculated as $(MAE_0 - MAE_{10})/(MAE_0 - MAE_{1000})$. (c, d) Absolute prior sensitivity S(n), defined as the MAE range over one-hot, k-hot and continuous inputs for H, O and F at n = 0, 2, 4 and 10, for ALIGNN and CGCNN, respectively. Error bars show standard deviations over three replicates; replicate and seed assignments are given in Supplementary Table S2.

The fixed-prior error response is accompanied by convergence among different initial representations. For H, O and F, the MAE range over one-hot, k-hot and continuous priors decreases from roughly 1.25, 2.50 and 1.82 eV/atom at n = 0 to 0.065, 0.046 and 0.096 eV/atom at n = 10. The two observations are complementary: adding target support reduces error without changing the k-hot input, and it simultaneously reduces the performance consequences of choosing among the tested priors.

To test whether this convergence is specific to ALIGNN, we repeat a targeted comparison with CGCNN for H, O and F. Across one-hot, k-hot and continuous inputs at n = 0, 2, 4 and 10, CGCNN shows the same qualitative trend: representation sensitivity is greatest without target support and decreases after target-containing structures are introduced. The detailed perturbation and adaptation controls remain specific to ALIGNN.

Together, the fixed-prior response and cross-architecture convergence identify target support as a variable that changes how strongly prediction depends on the initial element representation. We next examine both sides of this transition: the representation-induced sharing available without target labels and the learning enabled by target-containing environments.

### Representation-induced sharing shapes zero-shot transfer

Matched perturbations support the interpretation that zero-shot effects follow the element-sharing relations induced by the representation rather than input dimension or numerical form alone. Reassigning intact feature vectors among elements changes LOEO errors, whereas form-matched random inputs do not reproduce chemically organized priors; keep/drop controls further show that individual attributes can introduce either useful or harmful sharing. Full control definitions are provided in Methods; quantitative comparisons and feature-level results are provided in Supplementary Fig. S2 and the accompanying discussion.

### Reorganizing label-free element sharing changes zero-shot transfer

The perturbation results suggest that zero-shot performance depends on how the element representation organizes relations between the held-out element and elements observed during training. K-hot features encode such relations through hard bins: elements within the same bin share an active feature, whereas chemically similar elements separated by a bin boundary do not. Supplementary Haar[32,33] smoothing experiments show that weakening these boundaries can improve selected tasks, but the gains are not consistent across elements. This suggests that the relevant issue is not smoothing alone, but whether the induced element-similarity structure reflects chemically meaningful relations.

We therefore construct a label-free similarity-graph[34,35] prior from physicochemical attributes and periodic-table adjacency. The resulting graph coordinates provide a continuous element representation derived without target-property labels and replace the manually defined bin structure of k-hot features. This experiment tests whether reorganizing label-free chemical information can reduce zero-shot errors when no target-containing training structures are available.

The graph construction and preprocessing procedures are described in Methods under 'Element priors and similarity-graph coordinates.' Figure 4 compares the similarity-graph prior with k-hot and continuous representations across the six zero-shot LOEO tasks.

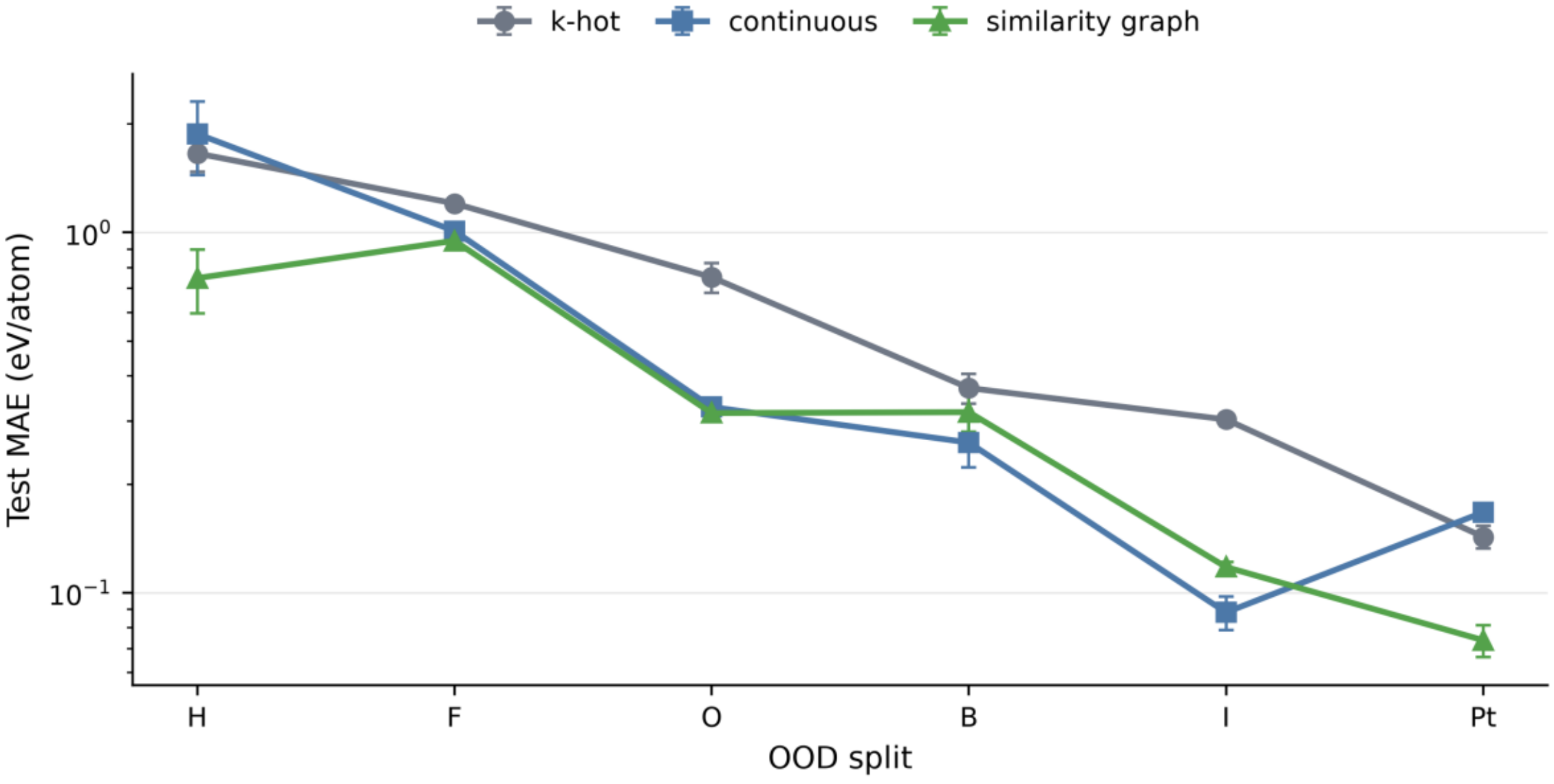

Fig. 4 | Zero-shot LOEO test error for k-hot, continuous and similarity-graph priors. Points show mean test MAE (eV/atom) for Leave-H/F/O/B/I/Pt-Out, and error bars show standard deviations over three training seeds.

The similarity-graph prior yields a lower worst-split error than k-hot and continuous across the six LOEO tasks. It has lower MAE than k-hot for each of the three seeds in all six tasks. Relative to continuous features, it sacrifices a small amount of accuracy on I and B but avoids the large error on H. Its highest error occurs in Leave-F-Out, where it remains lower than k-hot and continuous by 0.255 and 0.061 eV/atom, respectively.

The similarity-graph prior is a label-free alternative designed to test whether a different element-similarity structure improves zero-shot prediction, rather than a proposed universal atomic embedding. Its selected improvements show that static priors can still improve zero-shot prediction when target-property labels are unavailable. The target-support results then show that performance differences among static priors contract once target-support structures are introduced.

### Target support provides gains beyond output calibration and element-stem updates

The following diagnostics use a separate fixed training run and evaluation partition from the support-response experiments in Fig. 3; their zero-shot baselines therefore differ. We first test whether the gain from ten target-support structures can be explained by a simple correction of zero-shot predictions. Constant and affine residual calibration reduce the error for B, H, O and F, but neither approaches the performance of full retraining on the same support structures. For example, affine calibration reduces the MAE from 1.622 to 0.594 for H and from 1.504 to 0.634 for F, whereas retraining reaches 0.135 and 0.126, respectively. For I, neither calibration method improves substantially over the zero-shot baseline, while retraining reduces the MAE to 0.087. Thus, scalar and affine corrections explain only part of the improvement provided by target-containing training structures. The correction formulas and fitting protocol are described under "Target-support and adaptation analyses" in Methods.

We next compare target-containing compounds with the available pure-element reference structures. Elemental-only support produces substantially smaller gains than retraining on target-containing compounds across all six elements (Supplementary Fig. S3). Because the number and structural diversity of the elemental references are not matched to the random ten-structure support sets, this control does not isolate the effect of chemical environment from sample count or composition. Nevertheless, the results show that the available isolated-element structures do not reproduce the improvement obtained from target-containing compounds. The number of elemental-reference structures used for each element is reported in Supplementary Table S4.

Finally, we freeze the trainable element-projection layer, referred to here as the element stem, during retraining with ten target-support structures. The frozen-stem models remain close to full retraining for Pt, I, B, H and O, whereas the MAE for F increases from 0.126 to 0.229. Most of the observed improvement in five of the six tasks therefore persists without updating the initial element projection, while the larger degradation for F suggests a greater contribution from element-stem adaptation in this case. Taken together, the calibration, elemental-reference and frozen-stem controls show that the target-support gain cannot be attributed solely to output correction, isolated-element exposure or relearning of the initial element projection. The remaining improvement may arise from changes in message passing layers, the readout layer or their joint adaptation to target-containing environments.

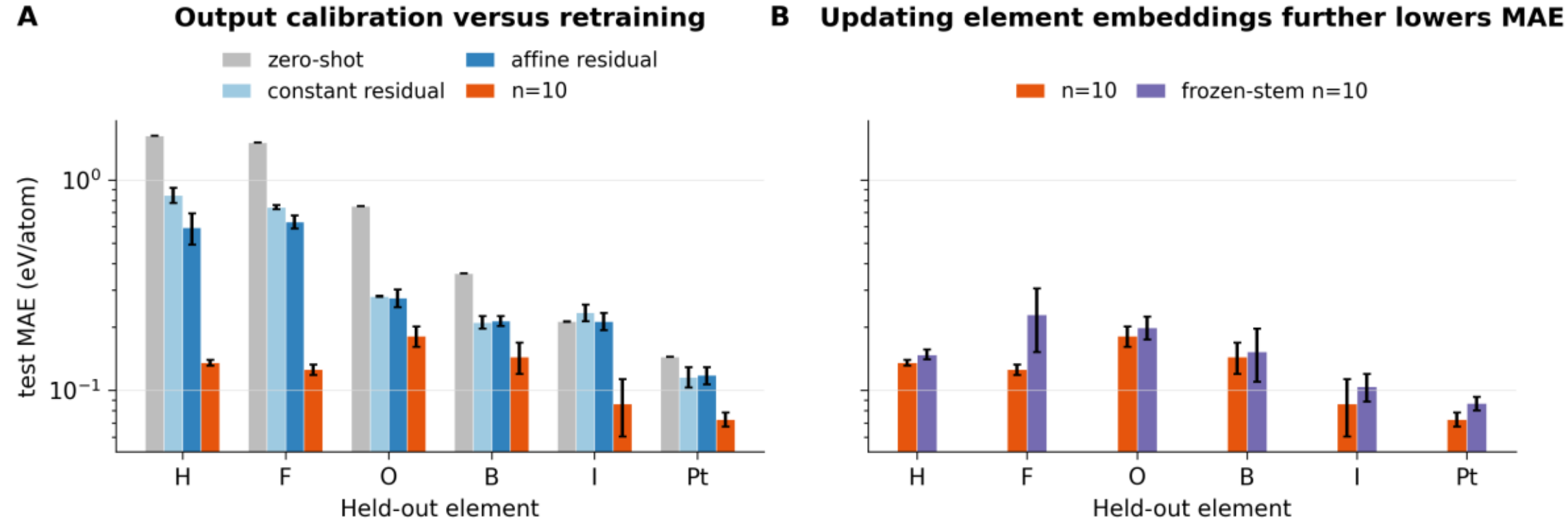

Fig. 5 | Output-calibration and element-stem update controls at n = 10 target-support structures. (a) Target-test MAE for zero-shot prediction, constant residual calibration, affine residual calibration and full retraining with ten randomly sampled target-containing structures (n = 10). (b) Target-test MAE for full n = 10 retraining and n = 10 retraining with the initial element-projection layer frozen (frozen-stem n = 10). Error bars show standard deviations over three support-sampling seeds. The calibration experiment uses a separate fixed training run and evaluation partition from Fig. 3. Calibration definitions are given in Methods; elemental-only support is reported separately in Supplementary Fig. S3 and Table S4.

## Element-prior sensitivity decreases when target chemistry is represented

Temporal drift and Leave-Triclinic-Out retain target chemistry in training and show substantially narrower variation across element priors than zero-shot LOEO (Fig. 6). Leave-Triclinic-Out shows substantially lower prior sensitivity than the LOEO tasks. Temporal drift is also less sensitive than the most prior-dependent LOEO splits, although its sensitivity remains comparable to several individual held-out-element tasks. These controls are therefore consistent with, but do not by themselves establish, a weaker role for element priors when target chemistry remains represented.

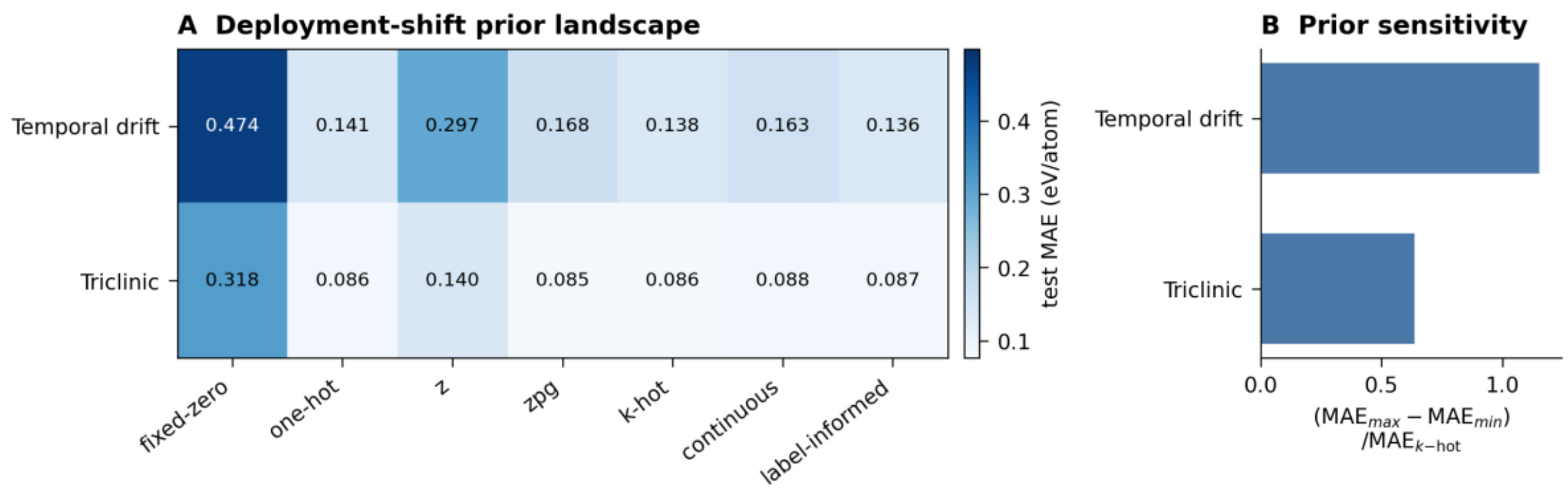

Fig. 6 | Element-prior landscape in deployment-shift controls. (a) Mean test MAE (eV/atom) over three training seeds for temporal drift and Leave-Triclinic-Out using fixed-zero, one-hot, z, zpg, k-hot, continuous and label-informed inputs. Cell labels report the mean MAE. (b) Normalized prior sensitivity over one-hot, z, zpg, k-hot and continuous inputs.

# DISCUSSION

This study identifies target support as a variable that changes which information source limits chemical transfer. Without target-containing training structures, prediction depends strongly on the relations and sharing paths supplied by the element representation. As target support is added, errors fall rapidly and performance differences among the tested priors contract, indicating reduced dependence on the initial input organization.

Under zero-target support, descriptor richness alone does not determine transfer. Matched perturbations support a role for element-feature correspondence beyond input dimension and numerical form, while the similarity-graph intervention shows that reorganizing label-free element relations can reduce selected zero-shot errors. These results define a role for prior design when target labels are unavailable without establishing a universally superior embedding.

The support-response curves identify the other side of this transition. Most of the measured improvement occurs within the first ten structures, and scalar or affine output corrections recover only part of that gain. Elemental-only support likewise fails to reproduce the recovery from target-containing compounds. Freezing the element stem retains most recovery in five of six splits, indicating that target support can usually be exploited without rewriting the initial element representation. Target support therefore changes how the existing representation is used in target-containing environments rather than merely correcting a zero-shot output scale or exposing the model to target identity.

Taken together, the support axis connects two transfer regimes that are usually entangled in conventional evaluation. Its zero-target-support endpoint exposes the indirect route supplied by static element relations; incremental support reveals how rapidly environment-conditioned supervision reduces that constraint. The front-loaded response is consistent with a rapid reduction in reliance on predefined element relations once labeled target-containing environments become available. It is more than a simple output-calibration effect, although the present controls do not fully identify which downstream network components account for the remaining gain.

Prior design and target-data acquisition consequently address different information regimes. When target-property labels are unavailable, a chemically constrained label-free prior can reduce zero-shot risk. When labels can be obtained, even limited target-containing data can provide a more direct route to improvement. In settings where target chemistry is already represented, as in the temporal and triclinic controls, other limitations such as structural coverage and data quality can become more prominent.

# METHODS

### Dataset and evaluation design

We used Materials Project 2021 crystal structures to predict formation energy per atom. The primary evaluation comprised leave-one-element-out (LOEO) splits for Pt, I, B, H and O and a Leave-Column-8-Out split containing F, Cl and Br. For brevity, this latter split is referred to as Leave-F-Out. At zero target support, every structure containing the held-out element or column was removed from training, while the original validation split remained unchanged. Temporal-drift and Leave-Triclinic-Out splits served as boundary controls in which target chemistry remained represented.

For every $n > 0$ condition, structures are sampled from the original target-element test split and moved to training. The validation split remains unchanged. Within each sampling seed, one random permutation of candidate test indices is generated and larger support sets use prefixes of that permutation; support sets are therefore nested within a seed. The remaining test structures form the target-element evaluation set.

The ALIGNN k-hot support curves used $n = 2, 4, 10, 100$ and $1000$ target-containing structures. Prior-sensitivity analyses used $n = 2, 4$ and $10$ for H, O and F, and the CGCNN replication used $n = 0, 2, 4$ and $10$. Experiment-specific targets, priors, support sizes and replicate assignments are summarized in Supplementary Table S2.

### Models and training

ALIGNN was used for the primary analyses, and a targeted CGCNN replication covered H, O and F through ten target-support structures. Crystal graphs for both architectures used a 7.0 Angstrom cutoff. ALIGNN used a neighbor limit of 20, two ALIGNN layers followed by two graph-convolution layers, a node hidden dimension of 256, 16 radial-basis bins over 0-8 Angstrom and 40 angular-basis bins over -1 to 1. CGCNN used a neighbor limit of 12, three convolution layers, a node hidden dimension of 128 and 41 radial-basis bins over 0-8 Angstrom. Within each architecture, element inputs passed through a trainable projection and all remaining settings were held fixed across prior comparisons.

Models were trained for 25 epochs without early stopping using AdamW, a maximum learning rate of 0.001, weight decay of 0.00001 and a OneCycleLR schedule (pct_start = 0.3, div_factor = 25 and final_div_factor = 10000). The objective was mean-squared error, with batch sizes of 16 for ALIGNN and 64 for CGCNN. Training used 16-bit mixed precision on GPU, while model outputs and loss evaluation used float32. The checkpoint with the lowest validation RMSE was selected. Target means and sample standard deviations were recomputed from the current training split in every run, and predictions were returned to physical units before MSE, RMSE and MAE were evaluated.

### Element priors and similarity-graph coordinates

Element lookup vectors were defined over atomic numbers 1-94 and supplied through the same trainable element-projection layer. Fixed-zero assigned the same zero vector to every element; one-hot encoded 94 categorical identities; z used atomic number; zpg used atomic number, period and group; k-hot used the 92-dimensional binned physicochemical descriptor adopted by CGCNN; and continuous used nine numerical attributes. One-hot and k-hot inputs were not standardized. Positive ionization-energy and atomic-volume values were log transformed for the continuous input, without column-wise z-scoring. The label-informed reference instead initialized the element-projection weights from a random-split ALIGNN model and froze these transferred weights while training the subsequent interaction and readout layers on each LOEO task.

The similarity graph is constructed once over the 94 elements. The nine continuous attributes are group, period, electronegativity, covalent radius, valence, ionization energy, electron affinity, block and atomic volume. After the two positive-valued log transforms, each column is population z-scored. Separately standardized period and group are appended, yielding the 11-dimensional coordinates $x_e$ used for distance calculation. Pairwise weights are

$$K_{ij} = exp\left(-\frac{\|x_i - x_i\|^2}{h}\right)$$

where h is the median nonzero squared distance.

Periodic-table adjacency $C_{ij}$ is one for neighboring groups within a period or neighboring periods within a group. The graph is

$$A = K + C$$

with zero diagonal and no self-loops.

The graph is symmetrically normalized as

$$D^{-\frac{1}{2}} A D^{-\frac{1}{2}}$$

After eigendecomposition, the leading trivial component is discarded and the next 16 components are retained. Coordinates are

$$\psi_k(e) = \lambda_k \varphi_k(e)$$

Negative eigenvalues are clipped to zero before scaling, and all splits and seeds use the same graph. Component signs are fixed by making the largest-magnitude coordinate in each component positive. The 16-dimensional coordinates contain no formation-energy labels; only the subsequent projection layer is trainable.

## Haar smoothing diagnostic

Haar smoothing is used only as a diagnostic of the resolution of element sharing in the k-hot representation. Within each predefined k-hot segment, the feature vector is padded where necessary, transformed in an orthonormal Haar basis and truncated back to the original support. The coarse variant retains low-frequency within-segment contrasts, whereas the multiresolution variant adds higher-frequency components. Transformed features are normalized before being passed to ALIGNN.

## Continuous monotone-wrap control

For continuous attribute $j$, let $m_j$ be the median over elements and sample a positive scale $a_j$ from $U(0.5, 2.0)$. The transformed value is

$$u_{e_j} = \sinh^{-1}\left[a_j\left(x_{e_j} - m_j\right)\right]$$

Each transformed column is then linearly rescaled to recover the original column mean and population standard deviation. The transformation is monotone increasing and therefore preserves element ranking within every attribute while changing pairwise spacing. All runs use atom-feature transform seed 0.

The attribute-specific scales are listed in Supplementary Table S3.

## Zero-shot perturbation controls

These perturbations test a specific alternative to the sharing interpretation: zero-shot errors might depend only on vector dimension, sparsity or numerical scale, rather than on which feature vector is assigned to each element. We therefore alter one property of the input at a time while keeping the model and LOEO split fixed.

Correspondence controls reassign intact element vectors by row permutation; for k-hot, blockwise row permutation performs this reassignment within descriptor segments, whereas for zpg the Z and PG permutations separately disrupt atomic-number or period/group correspondence. Random-match controls replace chemical features with random inputs matched in broad dimension, sparsity or numerical distribution. Continuous controls retain or remove feature groups, delete individual attributes, or apply a monotone warp that preserves element ranking within each attribute while changing pairwise spacing.

For the grouped continuous controls, we define four feature subsets: Periodic (group, period and block), Physchem (electronegativity, covalent radius, valence, ionization energy, electron affinity and atomic volume), Electronic (electronegativity, valence, ionization energy and electron affinity) and Size (covalent radius and atomic volume). These subsets are retained or removed as specified in the corresponding comparison. Under a purely form-based explanation, correspondence changes should have little systematic effect. Under the sharing interpretation, changing the element-feature assignment should change transfer even when broad input form is preserved.

## Target-support and adaptation analyses

Support-response analyses used the support levels specified above, with prior-sensitivity comparisons focused on one-hot, k-hot and continuous inputs for H, O and F. Full-retraining, frozen-stem and calibration comparisons used matched support structures and fixed, disjoint evaluation subsets; MAE is reported in eV/atom.

Frozen-stem retraining uses the same $n = 10$ support samples, evaluation subsets and sampling seeds as full retraining, with training seed fixed at 42. Residual calibration fits constant and affine corrections to the same support structures without retraining the GNN. Elemental-reference structures are limited by the available candidate pool and are not count- or environment-matched to the random $n = 10$ target-support sets.

Output calibration started from zero-shot predictions and fitted two corrections on the same ten target-support structures used for the $n = 10$ retraining comparison. Constant residual calibration fitted a single offset

$$y_{cal} = y_{pred} + b$$

whereas affine residual calibration fitted a slope and an offset

$$y_{cal} = ay_{pred} + b$$

Neither correction updated the element projection, message-passing layers or readout. The fitted correction was applied to the disjoint target-element evaluation set. Frozen-stem retraining instead updated the remaining network parameters while holding the initial element-projection layer fixed. Elemental-reference sample counts and selection conventions are reported in Supplementary Table S4. These controls are compared in Fig. 5 and Supplementary Fig. S3.

### Statistics and reproducibility

The zero-shot ALIGNN prior landscape and the n = 0 CGCNN baselines were each trained with three independent initialization seeds (0, 42 and 123). Target-support, prior-sensitivity, frozen-stem and calibration analyses used training seed 42 and three independently sampled support permutations (sampling seeds 13, 29 and 43). Within each sampling seed, support sets were nested across n and evaluation structures were disjoint from support. Reported values are arithmetic means, and error bars or ± values denote sample standard deviations across the three training or sampling replicates, as appropriate. Here, n is the number of target-support structures. No null-hypothesis significance tests were performed; comparisons are descriptive and use MAE in eV/atom. The exact replicate structure and the smaller complete candidate pools used for selected elemental-reference controls are reported in Supplementary Tables S2 and S4.

# DATA AVAILABILITY

The Materials Project data used in this study are publicly available from the Materials Project. Processed split files, element-feature tables, target-support samples and generated benchmark results will be made available upon publication.

# CODE AVAILABILITY

The code used to train models, generate OOD splits, construct element priors, sample target-support sets and reproduce the analyses will be made available upon publication.

# ACKNOWLEDGEMENTS

The authors acknowledge institutional support from KAUST. For computer time, this research used Shaheen III and Ibex managed by the Supercomputing Core Laboratory at King Abdullah University of Science & Technology (KAUST) in Thuwal, Saudi Arabia.

# AUTHOR CONTRIBUTIONS

R.Z. and K.L. designed experiments, R.Z. performed the experiments, analyzed the data, prepared the figures and wrote the manuscript. K.L. conceived and supervised the study, interpreted the results and revised the manuscript.

# COMPETING INTERESTS

The authors declare no competing interests.

# ADDITIONAL INFORMATION

Supplementary information will be provided with the online version of the paper. Correspondence and requests for materials should be addressed to the corresponding author.